\documentclass[aps,groupedaddress,superscriptaddress,amsmath,amssymb,twocolumn,pra]{revtex4-1}
\usepackage{bm}
\usepackage[dvips]{graphicx}
\usepackage{wrapfig,subfigure}
\usepackage{stmaryrd}
\usepackage{color}
\usepackage[normalem]{ulem}
\usepackage{enumerate}
\usepackage{epstopdf}

\def\kb{{\bf k}}
\def\sgn{{\rm sgn}}

\def\fp{f_{\rm p}}
\def\mup{\mu_{\rm p}}
\def\ep{\varepsilon}
\def\ka{\varkappa}

\newcommand{\sumprime}[1]{\sum_{#1}{\vphantom{\sum}}^{\!\!\prime}}

\def\varkab{{\bm \varkappa}}

\begin{document}
\author{M. I. Dykman}
\affiliation{Department of Physics and Astronomy, Michigan State
  University, East Lansing, Michigan 48824, USA}
\author{Christoph Bruder}
\affiliation{Department of Physics, University of Basel,
  Klingelbergstrasse 82, CH-4056 Basel, Switzerland}
\author{Niels L\"orch}
\affiliation{Department of Physics, University of Basel,
  Klingelbergstrasse 82, CH-4056 Basel, Switzerland}
\author{Yaxing Zhang}
\affiliation{Department of Physics, Yale University, New Haven, Connecticut 06511, USA}

\title{Interaction-induced time-symmetry breaking in driven quantum oscillators}

\date{\today}
\begin{abstract}
We study parametrically driven quantum oscillators and show that,
even for weak coupling between the oscillators, they can exhibit
various many-body states with broken time-translation symmetry. In
the quantum-coherent regime, the symmetry breaking occurs via a
nonequilibrium quantum phase transition.
For dissipative oscillators, the main effect of the
weak coupling is to make the switching rate of 
an oscillator between its
period-2 states dependent on the states of 
other oscillators. This
allows mapping the oscillators onto a system of coupled spins. Away
from the bifurcation parameter values where the period-2 states
emerge, the stationary state corresponds to having a microscopic
current in the spin system, in the presence of disorder. In the
vicinity of the bifurcation point or for identical oscillators, 
the stationary state can be mapped on that of the Ising model 
with an effective temperature
$\propto \hbar$, for low temperature. 
Closer to the bifurcation point the coupling can not be considered weak 
and the system maps onto coupled overdamped Brownian
particles performing quantum diffusion in a potential landscape.
\end{abstract}
\maketitle

\section{Introduction}

Time-symmetry breaking in periodically modulated quantum systems,
often called a ``time crystal" effect \cite{Wilczek2012}, has been
attracting much attention recently. One of the most challenging
problems in this rapidly developing area is the understanding of the
interplay of interaction, disorder, and dissipation
\cite{Khemani2016,Else2016,Keyserlingk2016,Yao2017,Zhang2017,Choi2017,Rovny2018,Berdanier2018}. In
particular, disorder helps preventing heating of the system by a
periodic drive in the coherent regime
\cite{Abanin2016,Bordia2017,Abanin2017a}.  However, the dependence of
the lifetime of the broken-symmetry state on the disorder strength is
not known generally and is likely to be model-dependent. Disorder
should not be necessary in the presence of dissipation. An example is
the observation of an interaction-induced breaking of the
time-translation symmetry in a dissipative classical cold-atom system
\cite{Kim2006}. A microscopic theory mapped the effect onto a phase
transition in an all-to-all coupled Ising system, and the measured
critical exponents were in agreement with this mapping \cite{Heo2010}.

\begin{figure}[h]
\includegraphics[scale=0.2]{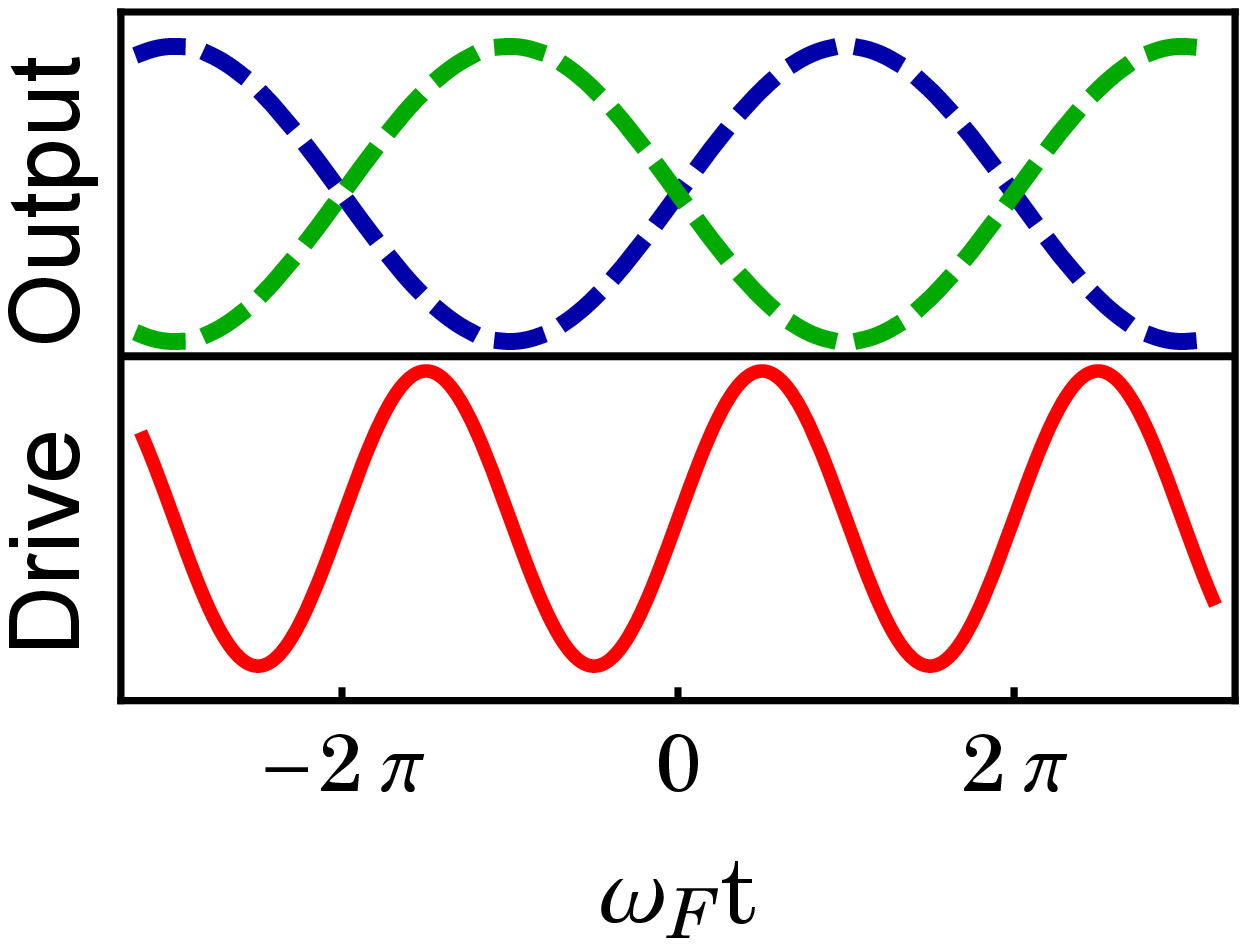}\hfill
\includegraphics[scale=0.35]{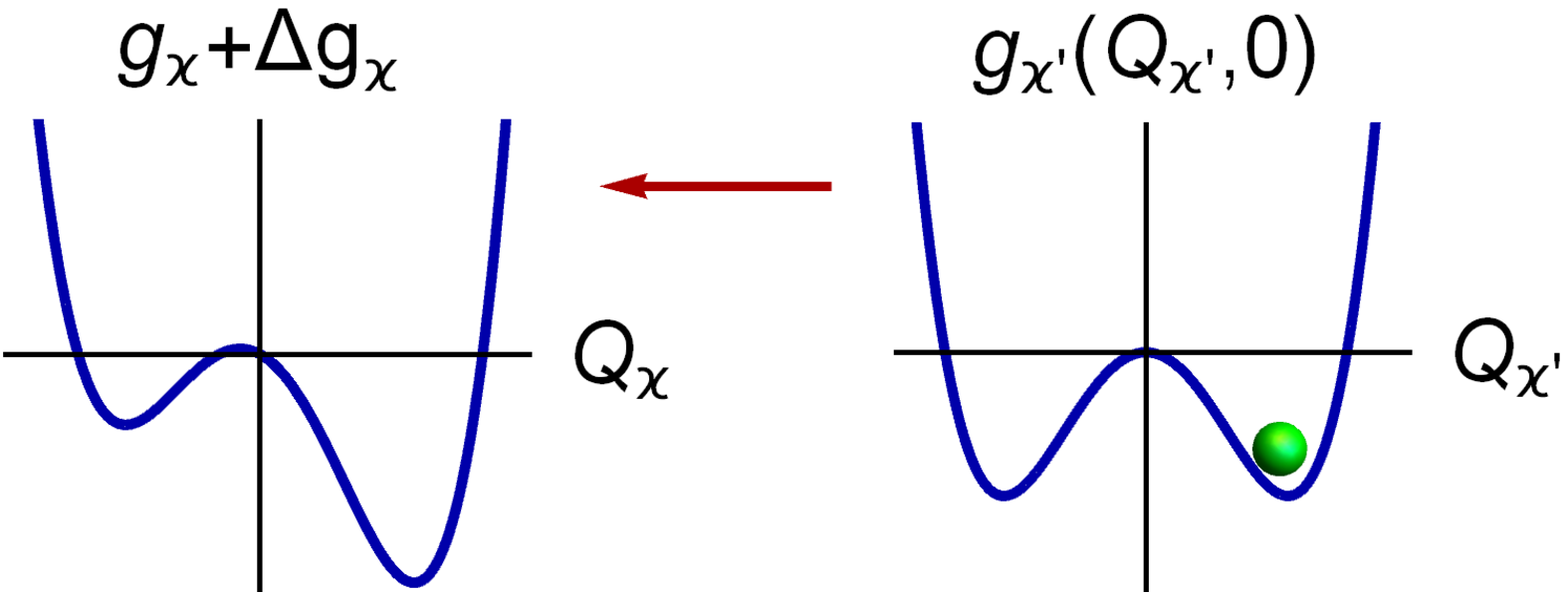}
\caption{Left panel: Period doubling of a weakly nonlinear classical
  dissipative oscillator whose eigenfrequency $\omega_0$ is modulated
  at frequency $\omega_F\approx 2\omega_0$. Right two panels: The
  effective double-well Hamiltonian of a parametric oscillator as a
  function of its coordinate in the rotating frame. The right panel
  refers to an isolated oscillator, where the Hamiltonian
  $g_{\ka'}(Q_{\ka'},P_{\ka'})$ is symmetric. The central panel shows
  how the Hamiltonian $g_\ka(Q_\ka,0)$ is modified by an extra term
  $\Delta g_\ka$ due to the coupling to an oscillator $\ka'$, which
  occupies a broken-symmetry state, as sketched in the right
  panel. The arrow indicates where the asymmetry is coming from.}
\label{fig:vibrations}
\end{figure}

Closely related to the problem of time-symmetry breaking is
\textit{computing} with parametric oscillators
\cite{Wang2013,McMahon2016,Inagaki2016a,Goto2016,Nigg2017,Puri2017,Goto2018}. A weakly
nonlinear classical dissipative oscillator displays period doubling
when its eigenfrequency $\omega_0$ is modulated at a frequency
$\omega_F$ close to $2\omega_0$ \cite{LL_Mechanics2004}. The emerging
period-2 states have opposite phases, see
Fig.~\ref{fig:vibrations}. If the system is in one of these states,
time-translation symmetry is broken, since the period of the motion is
$4\pi/\omega_F \approx 2\pi/\omega_0$ instead of $2\pi/\omega_F$.
These states can be associated with two states of a classical bit
\cite{Woo1971}, or two spin states. The spin analogy was studied in
recent numerical work for up to four coupled quantum parametric
oscillators and, for a number of parameter values, it was shown that
the system can be mapped onto an ``Ising machine" in the coherent
\cite{Goto2016} as well as the dissipative regime \cite{Nigg2017, Puri2017, Goto2018}.

In this paper we study the possibility and the nature of time-symmetry
breaking in a large system of coupled quantum parametric
oscillators. Our formulation applies in the presence of weak disorder,
but of primary interest to us are the broken-symmetry phases that
emerge in the coherent and dissipative regimes and the associated
phase transitions that occur even without disorder. The relevant
physical systems are microwave modes in superconducting cavities that
can be coupled into lattices with variable geometry
\cite{Underwood2016,Kollar2018,Kounalakis2018}, as well as coupled
vibrational modes in networks of nanomechanical resonators
\cite{Buks2002,Fon2017}. An advantageous feature of these systems is
the possibility to make them one- or two-dimensional, control the
coupling strength, and implement various coupling geometries which, at
least in the nanomechanical setting, are not limited to
nearest-neighbor coupling. We assume the coupling of the modes in
different resonators to be comparatively weak and the mode
eigenfrequencies to form a narrow band centered at a characteristic
eigenfrequency $\omega_0$.  In the absence of a periodic drive, the
spectrum of excitations is therefore also a narrow band centered near
$\omega_0$.

The effect of the coupling is significantly more complicated for
parametrically excited modes. To understand it, we note first that the
quantum dynamics of an isolated mode can be mapped onto the dynamics
of an auxiliary particle with a double-well quasienergy Hamiltonian
\cite{Marthaler2006}. The Hamiltonian is symmetric; the two minima
correspond to the opposite phases of the period-2 oscillations, see
the right panel in Fig.~\ref{fig:vibrations}. The mode can tunnel
between the minima, which leads to phase-flip transitions. However, if
the relaxation rate largely exceeds the tunnel splitting, the
interwell switching occurs via effectively ``overbarrier"
transitions. This happens even for $T=0$ due to quantum noise, which
invariably accompanies relaxation \cite{Marthaler2006}.  One can
associate the minima of the Hamiltonian of the $\ka$th mode with a
spin $\sigma_\ka =\pm 1$.  The switching rates
$W_{\sigma_\ka}\equiv W_{\sigma_\ka\to -\sigma_\ka}$ between the
minima are equal by symmetry.

If the modes are coupled and one of them is in a certain state (near one of
the minima of the quasienergy Hamiltonian, see
Fig.~\ref{fig:vibrations}), the symmetry of the effective Hamiltonian
for the mode coupled to it is broken. Then, for this mode, the
switching rates between the minima become different. Depending on the
sign of the coupling, the ``deeper" well corresponds to the
oscillators having the same (for the case of attractive coupling) or
the opposite (for the case of repulsive coupling) phase.

It is important that the rates $W_{\sigma_\ka}$ are much smaller than
the inverse $t_r^{-1}$ of the relaxation time. Therefore, when one of
the modes is switching, the modes coupled to it are most likely
localized in a certain minimum.  As we show, the change of the
switching rate $W_{\sigma_\ka}$ of mode $\ka$ due to its coupling to
modes $\ka'$ can be large even for weak coupling, with
$\log W_{\sigma_\ka}$ being linear in the coupling. This allows one to
map the problem onto the Ising model of coupled spins, for identical
oscillators.

The well-known properties of the Ising model imply that, for not too
weak attractive coupling, the most probable state of the many-mode
two-dimensional system is the broken-symmetry state with all
$\sigma_\ka$ equal, i.e., the phases of all oscillators being the
same. In this state, the symmetry with respect to time translation by
the drive period is broken. For the case of repulsive mode coupling,
the system of coupled modes maps onto the antiferromagnetically
coupled Ising model and can exhibit frustration, depending on the
geometry of the lattice and the structure of the coupling.

Explicit analytical expressions for the coupling parameters of the
Ising model can be obtained near the bifurcation point where the period-2
states of the individual modes emerge.
As the parameters approach this point, the effective coupling strength
increases. Ultimately, the system goes into the strong-coupling
regime, and the mapping on the Ising model becomes inadequate. The
modes strongly mix, and one can no longer think of an ensemble of
individual modes having their own double-well quasienergy Hamiltonian.

For strong coupling, the appropriate picture is that of a
multiple-well ``quasienergy landscape". This landscape has global
symmetry with respect to time translation $t\to t+2\pi/\omega_F$, but
each individual minimum does not have this symmetry. As a result,
there are many metastable broken-symmetry states. Quantum noise leads
to diffusion between these states, but the transitions between
different minima involve many modes and become exponentially slow. The
system effectively ``freezes" in one of them, and time-translation
symmetry is then broken.

\begin{figure}[h]
\includegraphics[scale=0.15]{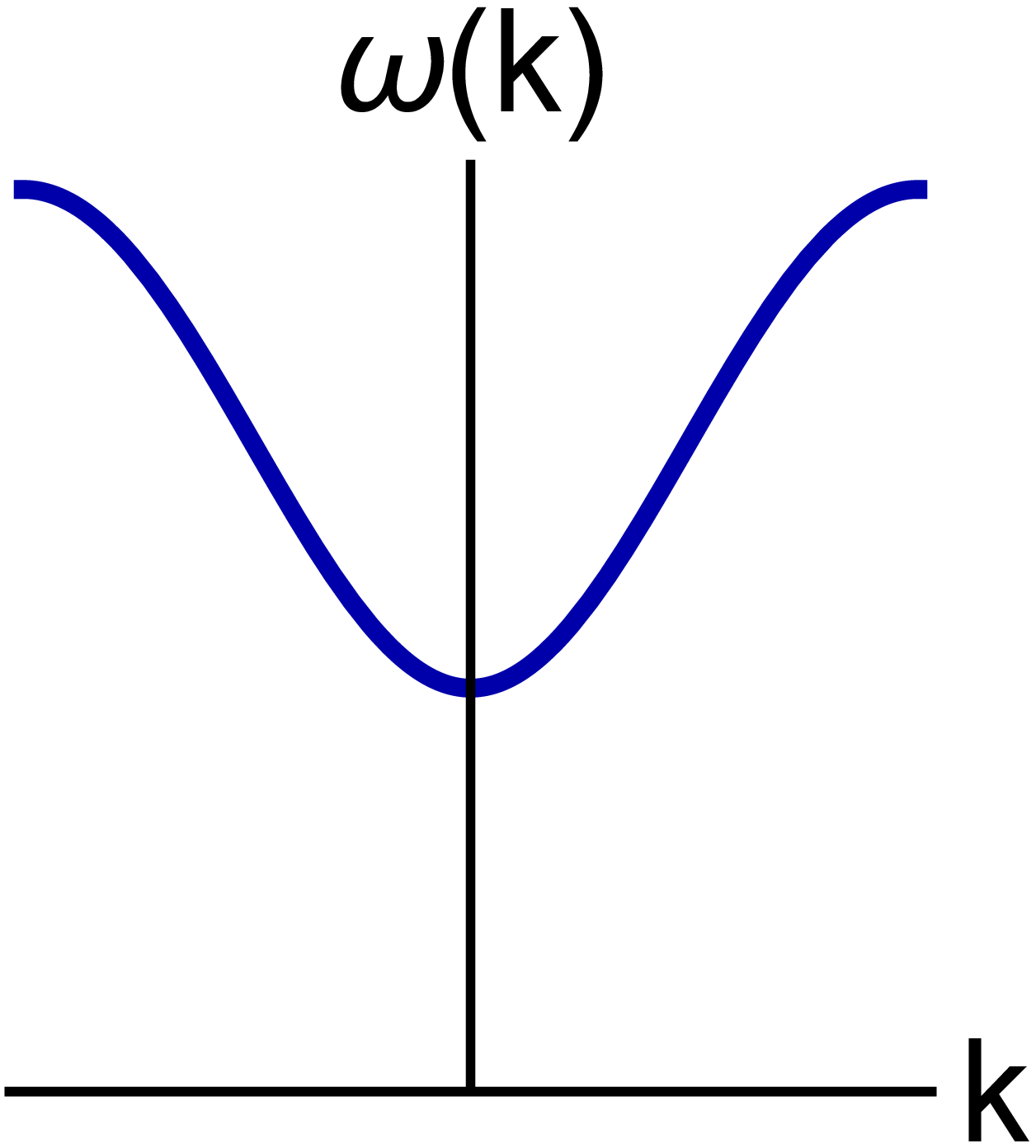}\hfill
\includegraphics[scale=0.15]{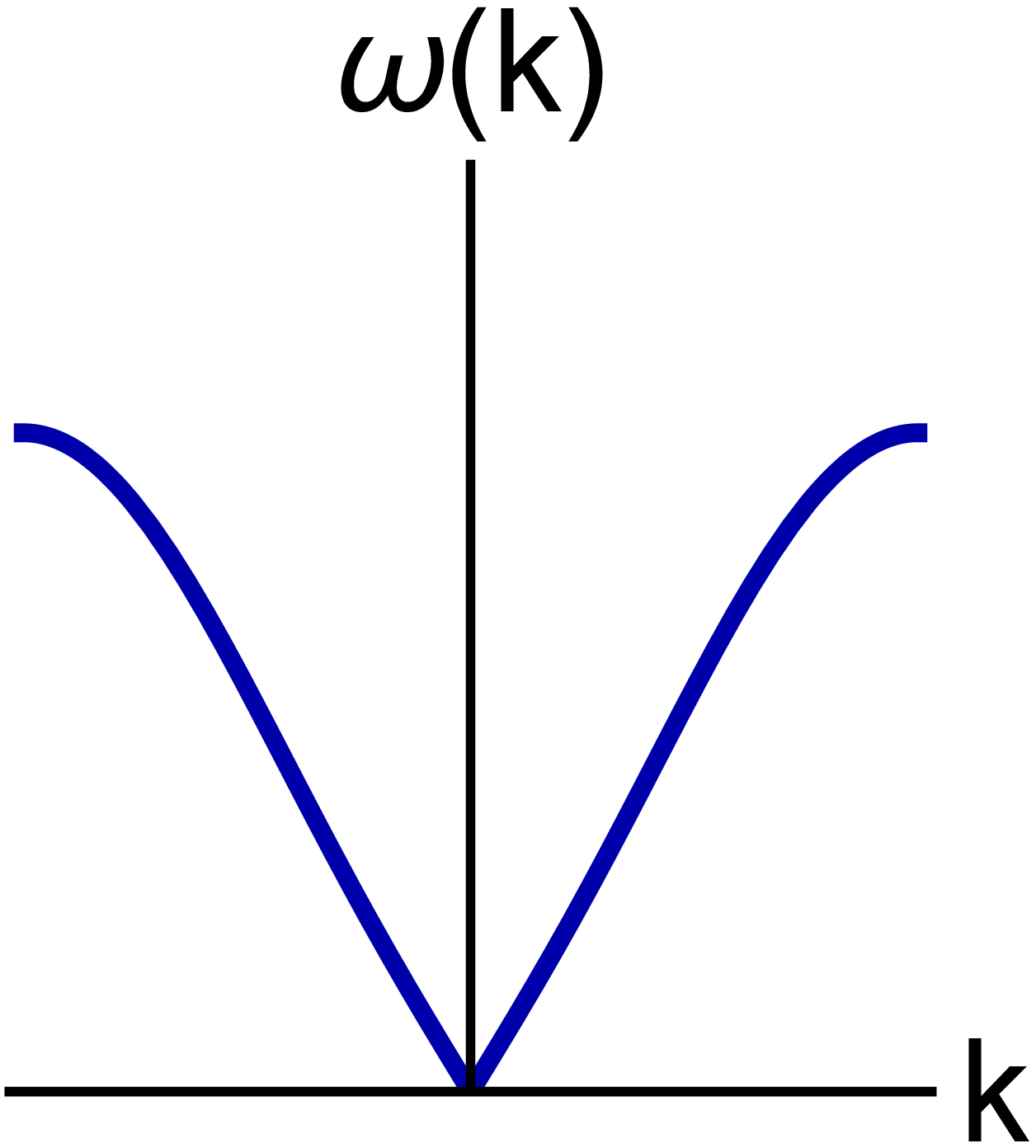}\hfill
\includegraphics[scale=0.15]{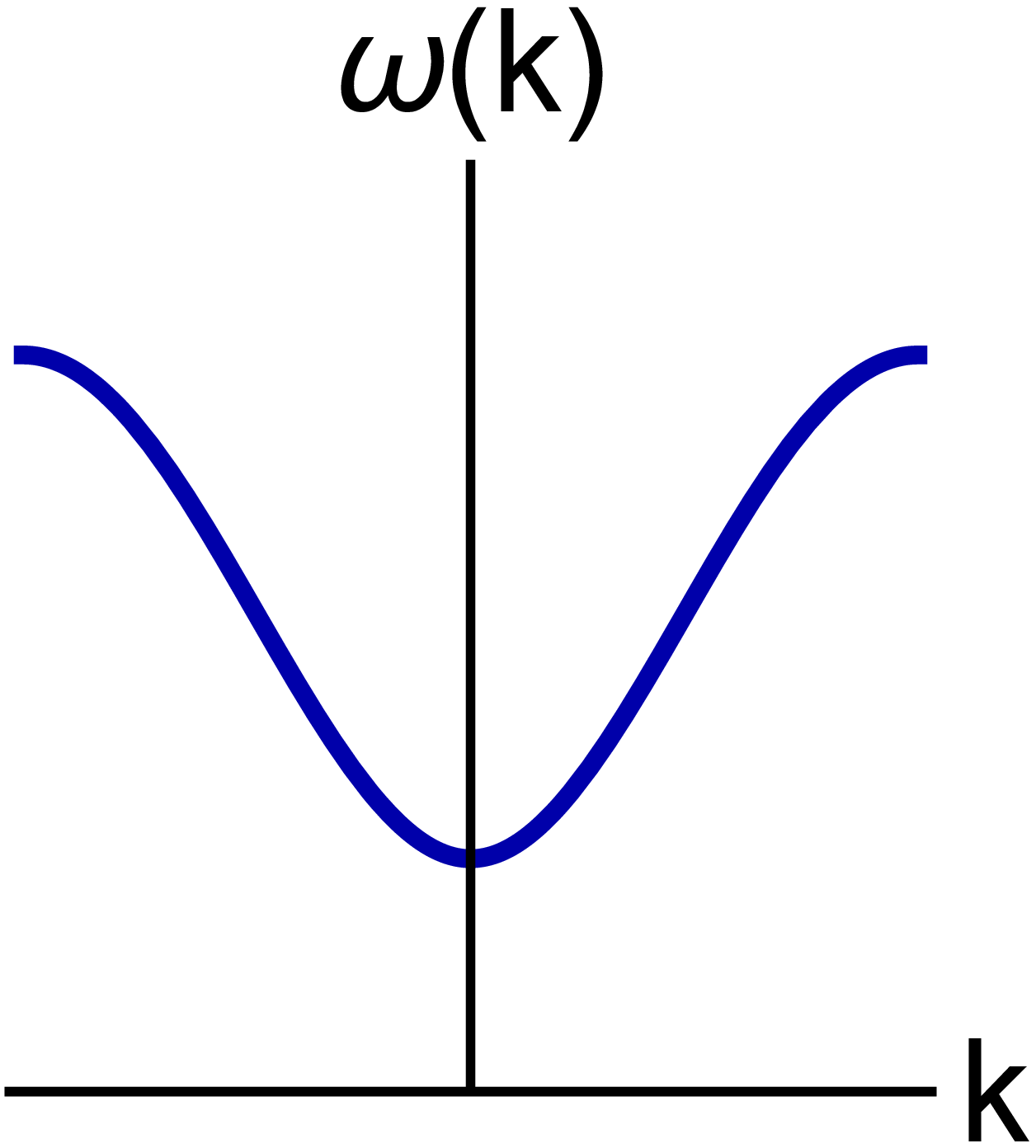}\hfill
\caption{Schematic view of the excitation spectrum $\omega(k)$ on
  going through the quantum phase transition. If the control parameter
$\mu$ is below (left panel) or above (right panel) the
critical value $\mu_{\rm QPT}$, the spectrum is gapped. For
$\mu=\mu_{\rm QPT}$, the spectrum becomes linear for $k\to 0$.}
\label{fig:QPT}
\end{figure}

In the quantum-coherent regime, a new phenomenon appears: instead of a
bifurcation point for each individual oscillator, the coupled modes
exhibit a nonequilibrium quantum phase transition (QPT). The control
parameter is the distance to the critical value of the drive frequency
$\omega_F=\omega_{\rm QPT}$, or to the critical value of the drive
amplitude. We will consider the case where there is no disorder. The
spectrum of excitations of the system can be naturally defined, if one
starts from below the QPT, where the modes are not excited and all of
them occupy the ground state. Here the spectrum is gapped, see
Fig.~\ref{fig:QPT}. The simplest way to picture this situation is to
think of the excitation spectrum of the coupled system in the absence
of the drive, with the excitation frequency downshifted by
$\omega_F/2$. The gap goes to zero at the phase transition point and
the dispersion law of the long-wavelength excitations becomes
linear. For attractive coupling between the modes, beyond the QPT the
system has a state where all modes vibrate in phase and the excitation
spectrum is again gapped. This state has broken time-translation
symmetry, a direct analog of the ferromagnetic state of an Ising chain
that goes through a QPT on varying the transverse magnetic field
\cite{Lieb1961,Sachdev1999}.

The paper is organzied as follows: 
In Sec.~\ref{sec:RWA} below we describe the Hamiltonian of the
system. In Sec.~\ref{sec:symmetry_breaking} we show how the problem of
weakly coupled parametric oscillators can be described in terms of the
exponentially strong modification of the rate of interstate switching
of an oscillator depending on the state of other oscillators. The
description is based on the notion of the logarithmic
susceptibility. It allows mapping the system onto an Ising system
provided there is detailed balance.  This is the case if the
oscillators are identical.  In Sec.~\ref{sec:LS} we describe how to
find the quantum logarithmic susceptibility for weak
damping. Section~\ref{sec:near_bifurcation_point} describes a theory
of the coupled oscillators near the threshold of parametric
excitation. It also describes a spin-glass type case where the system
can have many metastable states with broken time-translation
symmetry. Section~\ref{sec:QPT} describes a quantum time-symmetry
breaking transition in a spatially-periodic system of coupled oscillators.
Section~\ref{sec:conclusions} contains concluding remarks.

\section{The model}
\label{sec:RWA}

We consider a system of coupled quantum oscillators (modes). They are weakly nonlinear and are parametrically modulated. The Hamiltonian of the system is
\begin{align}
 \label{eq:Hamiltonian}
 H=H_0 + H_F + H_c\:,
\end{align}
where
\begin{align}
\label{eq:H0}
  H_0=\frac{1}{2}\sum_\ka  (p_\ka ^2 + \omega_\ka ^2 q_\ka ^2)
  + \frac{1}{4} \gamma \sum _\ka  q_\ka ^4\:.
\end{align}
Here, $\ka =1,2,\ldots,N$ enumerates the oscillators, $q_\ka $ and
$p_\ka $ are their coordinates and momenta, and $\omega_\ka $ are
their eigenfrequencies; we assume that the values of $\omega_\ka $ are
close to each other, $|\omega_\ka -\omega_0|\ll \omega_0$. The
parameter $\gamma$ characterizes the lowest-order nonlinearity that is
relevant for resonantly excited small-amplitude oscillations
\cite{LL_Mechanics2004}. In what follows we assume $\gamma >0$; an
extension to the case $\gamma <0$ is straightforward.

The Hamiltonian $H_F$ describes resonant parametric driving,
\begin{align}
\label{eq:driving}
H_F = \frac{1}{2}\sum_\ka  q_\ka ^2 F\cos\omega_Ft,\quad \omega_F \approx 2\omega_0\:,
\end{align}
and $H_c$ is the Hamiltonian of the coupling between the modes,
\begin{align}
\label{eq:coupling}   
H_c = -\frac{1}{2}\sum_{\ka \neq \ka'} \ep_{\ka \ka'} q_\ka q_{\ka'}\:, \quad |\ep_{\ka \ka'} |\ll \omega_0^2\:.
\end{align} 
This interaction corresponds to bilinear mode coupling and occurs,
e.g., in microwave cavity arrays and in systems of mechanical
nanoresonators \cite{Kollar2018,Buks2002,Fon2017}.  The oscillator
nonlinearity, the coupling, and the driving are assumed to be weak,
$\gamma\langle q_\ka ^2\rangle, |\ep_{\ka \ka'} |, |F| \ll
\omega_0^2$. In this case the motion of the oscillators corresponds to
vibrations at frequency $\approx \omega_F/2$ with amplitude and phase
that slowly vary on the time scale $1/\omega_F$. This motion can be
conveniently described in the rotating frame by introducing the ladder
operators $a_\ka ,a^\dagger_\ka $ of the $\ka$th oscillator, applying
a canonical transformation
$U(t)=\exp[-i(\omega_Ft/2) \sum_\ka a_\ka ^\dagger a_\ka ]$ and
switching to the scaled coordinates $Q_\ka $ and momenta $P_\ka $ that
slowly vary in time,
\begin{align}
\label{eq:dimensionless}
&U^\dagger(t)[q_\ka +(2i/\omega_F)p_\ka ]U(t)=-iC(Q_\ka +iP_\ka )
e^{-i\omega_Ft/2},\nonumber\\
&[P_\ka ,Q_{\ka'}]=-i\lambda \delta_{\ka \ka'} ,\qquad \lambda=3\hbar \gamma/\omega_F F.
\end{align}
Here, we set $C= (2F/3\gamma)^{1/2} =(2\hbar/\lambda\omega_F)^{1/2}$;
in Appendix \ref{sec:scaled_amplitude} we use a different scaling to
describe the quantum phase transition induced by the varying field
amplitude. The parameter $\lambda$ is the dimensionless Planck
constant in the rotating frame; we note that, in terms of the scaled
variables $Q_\ka,P_\ka$, the lowering operator is
$a_\ka =(2\lambda)^{-1/2}(Q_\ka +iP_\ka )$.

We assume that the scaled Planck constant is small, $\lambda \ll
1$. This means that the dynamics in the rotating frame is
semiclassical and the states of parametrically excited period-2
oscillations of the individual mode overlap only weakly.

\subsection{The rotating wave approximation}

For weak nonlinearity and weak mode coupling, the resonant dynamics of
the coupled modes can be conveniently described in the rotating wave
approximation (RWA).  In this approximation the canonically
transformed Hamiltonian of the system becomes
\begin{align}
\label{eq:scaled_Hamiltonian}
&U^\dagger H U-i\hbar U^\dagger\dot U  \approx (3\gamma C^4/8) {\mathbb G}\:,\nonumber\\
&{\mathbb G}=\sum_\ka  g_\ka (Q_\ka , P_\ka ) +g_c\:.
\end{align}
Here, ${\mathbb G}$ is the scaled RWA Hamiltonian of the system. It is
the sum of the scaled RWA Hamiltonians
$g_\ka (Q_\ka ,P_\ka )\equiv g_\ka (Q_\ka ,-i\lambda \partial_{Q_\ka
})$ of the individual oscillators and the coupling term $g_c$. The
individual Hamiltonians $g_\ka$ depend on a single parameter
$\mu_\ka $ and can be expressed as \cite{Marthaler2006}
\begin{align}
\label{eq:g_period_2}
&g_\ka (Q_\ka ,P_\ka ) = \frac{1}{4}(P_\ka ^2+Q_\ka ^2-\mu_\ka )^2 + \frac{1}{2}(P_\ka ^2 -Q_\ka ^2)-\frac{\mu_\ka ^2}{4}\:,\nonumber\\
&\mu_\ka  = 2\delta\omega_\ka\omega_F/F, \qquad \delta\omega_\ka = \frac{1}{2}\omega_F - \omega_\ka\:.
\end{align}
The parameter $\mu_\ka$ is determined by the ratio of two small
parameters, the detuning $\delta\omega_\ka$ of half the drive
frequency from the mode eigenfrequency and the scaled drive amplitude $F/\omega_F$.
For $\mu_\ka <-1$, $g_\ka(Q_\ka ,P_\ka)$ has a single minimum at
$Q_\ka =P_\ka =0$. This minimum corresponds to the equilibrium
position of the oscillator in the laboratory frame. As $\mu_\ka$
increases beyond $-1$, the point $Q_\ka =P_\ka =0$ becomes first a
saddle point, and then, for $\mu_\ka >1$, a local maximum of $g_\ka
$. In addition, for $\mu_\ka >-1$, the function $g_\ka $ has two
symmetrically located minima at
$P_\ka =0, Q_\ka =\pm (\mu_\ka +1)^{1/2}$. They can be seen in the
right panel of Fig.~\ref{fig:vibrations}, which shows $g_\ka $ for
$P_\ka =0$. Classically, in the presence of weak dissipation these
minima become stable states. They correspond to two states of period-2
oscillations with opposite phases. We enumerate them by
$\sigma_\ka=\pm 1$; for concreteness, we set $\sigma_\ka=1$ to
correspond to the minimum of $g_\ka$ with $Q_\ka >0$.

The term $g_c$ in Eq.~(\ref{eq:scaled_Hamiltonian}) describes the
coupling Hamiltonian in the rotating frame,
\begin{align}
\label{eq:coupling_RWA}
&g_c=
-\frac{1}{2}\sum_{\ka \neq \ka'} V_{\ka \ka'} (Q_\ka Q_{\ka'} + P_\ka P_{\ka'})\:,\nonumber\\
&V_{\ka \ka'}  = 2\ep_{\ka \ka'}/F\:. 
\end{align}
We note that the coupling in the coordinate channel in the lab frame
described by Eq.~(\ref{eq:coupling}) becomes symmetric with respect to
the coordinates and momenta in the rotating frame, in the RWA.

The effect of the coupling on the mode dynamics depends on the
relation between $|V_{\ka\ka'}|$ and the depth of the wells of the
functions $g_\ka$, see Fig.~\ref{fig:vibrations}. If
$|V_{\ka\ka'}|\ll 1$, the overall many-mode Hamiltonian
(\ref{eq:scaled_Hamiltonian}) is a set of double-well functions
$g_\ka$ slightly distorted by the coupling (unless $\mu_\ka - 1$ is
also small, see below). If, on the other hand, the coupling is
comparatively strong, the overall structure of the Hamiltonian
changes. We will not consider this case in the present paper.

\subsubsection{Symmetry arguments}

The RWA Hamiltonian ${\mathbb G}$ has inversion symmetry, both in the
coordinates, $\{Q_\ka \to -Q_\ka\}$, and in the momenta,
$\{ P_\ka \to -P_\ka\}$. This symmetry is a consequence of the
symmetry of the Hamiltonian $H$ with respect to time translation by
the driving period $t\to t+2\pi/\omega_F$. From
Eq.~(\ref{eq:dimensionless}), such a translation corresponds to
changing the signs of $\{Q_\ka,P_\ka\}$. Indeed, as a result of the
time translation, the unitary operator $U(t)$ in
Eq.~(\ref{eq:dimensionless}) becomes $U(t+2\pi/\omega_F) = U(t)N_2$,
where $N_2=\exp(-i\pi\sum_\ka a_\ka^\dagger a_\ka)$. The
time-translation operator $N_2$ flips the sign of the mode coordinates
and momenta, $N_2^\dagger q_\ka N_2 = -q_\ka$ and similarly for
$p_\ka$. In addition, $N_2$ commutes with ${\mathbb G}$. Therefore, as
in the case of a single oscillator \cite{Guo2013a,Zhang2017b}, the
eigenfunctions of ${\mathbb G}$ are the Floquet eigenfunctions of the
original time-periodic Hamiltonian $H$, and the eigenvalues of
${\mathbb G}$ are the RWA-quasienergies of the system scaled by the
factor $3\gamma C^4/8$.

The individual RWA Hamiltonians $g_\ka$ also have inversion symmetry,
cf. Fig.~\ref{fig:vibrations}. Therefore, generally, the intrawell
states of $g_\ka$ are tunnel-split into symmetric and antisymmetric
states.  For a small dimensionless Planck constant $\lambda$, this
splitting is exponentially small deep inside the wells and may be
equal to zero for certain $\mu_\ka$ \cite{Marthaler2007a,*Zhang2017a}.

\subsection{Quantum kinetic equation}

We now discuss the dissipative dynamics of the system of parametric
oscillators. To this end, we will assume that each oscillator is
coupled to its own thermal reservoir and that all reservoirs have the
same temperature. We will use the simplest model where the interaction
with the reservoirs is linear in $q_\ka,p_\ka$.  If the densities of
states of the reservoirs weighted with the coupling to the environment
are sufficiently smooth near $\omega_0$, the dynamics of the nonlinear
oscillators in ``slow time''
\begin{align}
\label{eq:slow_time}
  \tau \equiv tF/2\omega_F
\end{align}  
is Markovian. A derivation is a straightforward extension of the
derivation for a single nonlinear oscillator \cite{DK_review84} to the
case of coupled oscillators; the frequency renormalization is
incorporated into $\omega_\ka$. For simplicity, we will assume that
the decay rates of all the oscillators are the same: different decay
rates constitute a dissipative type of disorder and will not be
discussed in this paper. With these assumptions, the master equation
for the multi-oscillator density matrix $\rho$ reads
\begin{align}
\label{eq:master_full}
&\dot\rho \equiv \frac{d\rho}{d\tau}=\frac{i}{\lambda}\left[\rho,{\mathbb G}\right] + \kappa\sum_\ka {\cal D}[a_\ka ]\rho\:,\nonumber\\
&{\cal D}[a]\rho = - \left(\bar n +1\right)\left(a^{\dagger}a\rho - 2a\rho a^{\dagger} + \rho a^{\dagger}a\right)  \nonumber \\ 
&- \bar{n} \left( a a^{\dagger}\rho  -2a^{\dagger} \rho a + \rho a a^{\dagger}\right)\:.
\end{align}
Here, $\kappa$ is the dimensionless oscillator decay rate and $\bar n
= [\exp(\hbar\omega_F/2k_BT)-1]^{-1}$ is the oscillator Planck
  number.

Alternatively, and this will be used below, one can write down the quantum Langevin equations 
\begin{align}
\label{eq:full_Langevin}
 \dot Q_\ka  = -\kappa Q_\ka  +\partial_{P_\ka }{\mathbb G} + f_{Q_\ka}(\tau)\:, \nonumber\\
\dot P_\ka  = -\kappa P_\ka  -\partial_{Q_\ka } {\mathbb G} +  f_{P_\ka}(\tau)\:.
\end{align}
Here, $f_{Q_\ka}(\tau)$ and $f_{P_\ka}(\tau)$ are
$\delta$-correlated operators, 
 \begin{align}
 \label{eq:noise_correlators}
 &\langle  f_{Q_\ka}(\tau)  f_{Q_{\ka'}}(\tau')\rangle =  \langle  f_{P_\ka}(\tau)  f_{P_{\ka'}}(\tau')\rangle \nonumber\\
 &= 2D\delta(\tau - \tau') \delta_{\ka \ka'},\qquad D=\frac{1}{2}\lambda\kappa (2\bar n +1),
 \end{align}
and $\langle [ f_{Q_\ka}(\tau),  f_{P_{\ka'}}(\tau')]\rangle =
2i\lambda\kappa\delta(\tau - \tau')\delta_{\ka\ka'}$. For small
$\lambda$ the noise intensity $D$ is small.
Equation~(\ref{eq:full_Langevin}) is the Heisenberg version of
the master equation (\ref{eq:master_full}). The partial derivatives of
${\mathbb G}$ in Eq.~(\ref{eq:full_Langevin}) should be interpreted as
symmetrized expressions, for example,
$\partial_P (P^2+Q^2)^2 = 2P(P^2+Q^2) + 2(P^2+Q^2)P$.

\section{Time-symmetry breaking for weakly coupled modes}
\label{sec:symmetry_breaking}

In this section we show that even weak mode coupling can lead to a
collective breaking of the time-translation symmetry if the quantum
noise is sufficiently weak. The underlying mechanism is the
coupling-induced change of the rate of switching between the period-2
states of the oscillators.

The dissipative dynamics of an isolated parametric oscillator $\ka$ is
characterized by the dimensionless relaxation rate $\kappa$ and by the
dimensionless rate of switching
$W_{\sigma_\ka} \equiv W_{\sigma_\ka\to -\sigma_\ka}$ between the
wells $\sigma_\ka$ and $-\sigma_\ka$ of the RWA Hamiltonian $g_\ka$.
The switching rates are exponentially smaller than the relaxation
rate, i.e., $W_{\sigma_\ka}\ll \kappa$. The oscillator approaches one
of the minima of $g_\ka$ on a time scale $\sim\kappa^{-1}$.  It
performs quantum fluctuations about this minimum for a time much
longer than $\kappa^{-1}$, until ultimately it switches to the other
minimum.  We note that the notation $W_{\sigma_\ka}$ is a shortcut: it
refers to the rate $W_\ka(\sigma_\ka)$ for the $\ka$th oscillator to
switch from the well $\sigma_\ka$ to $-\sigma_\ka$.

If $\kappa$ largely exceeds the exponentially small tunnel splitting
of the intrawell states, the interwell switching occurs via
``overbarrier" transitions \cite{Marthaler2006}. Such transitions
result from quantum diffusion over the intrawell quasienergy states,
which brings the system from the bottom of the initially occupied well
of $g_\ka$ to the top of the barrier. This process is reminiscent of the
familiar thermally activated overbarrier transitions in classical
systems \cite{Kramers1940}, except that, for low temperatures, it is
induced by quantum fluctuations and is called quantum activation. The
physical cause of the diffusion over quasienergy states is that
quantum relaxation is invariably associated with noise. Relaxation
results from transitions between the states of the oscillator with
emission of excitations of the thermal reservoir, but these
transitions happen at random, and therefore they bring in noise. The
presence of this quantum noise is reflected in the noise terms in
Eq.~(\ref{eq:full_Langevin}).

For an isolated oscillator $\ka$, the rate of switching due to quantum
activation has the form
\[W_{\sigma_\ka}^{(0)} = \mathrm{const}\times \exp(-R_{\sigma_\ka}^{(0)}/\lambda).\] 
The parameter $R_{\sigma_\ka}^{(0)}$ is the quantum activation energy
of switching from state $\sigma_\ka$; note that the quantum noise
intensity $\lambda$ plays here a role analogous to temperature in the
expression for the rate of thermally activated switching.  By
symmetry, $R_{\sigma_\ka}^{(0)}$ is the same for the both states
$\sigma_\ka=\pm 1$.  Expressions for $R_{\sigma_\ka}^{(0)}$ have been
found in several important limiting cases
\cite{Marthaler2006,Dykman2007}.

\subsection{Symmetry lifting by an extra field at frequency $\omega_F/2$}
\label{subsec:extra_field}

Before analyzing the effect of the coupling of the modes, we consider
a simpler related problem, viz., the effect of a weak additional field
at frequency $\omega_F/2$ on the switching rate. Such a field is
described by the term $-F'\sum_\ka q_\ka\cos(\varphi_\ka+\omega_Ft/2)$
in the Hamiltonian. It breaks the time-translation symmetry
$t\to t+2\pi/\omega_F$. In the rotating frame, the effect of
the field $\propto F'$ on the mode dynamics is described by the term
\begin{align}
\label{eq:extra_field}
\Delta g_\ka(Q_\ka,P_\ka) = -f'(Q_\ka\sin\varphi_\ka + P_\ka\cos\varphi_\ka)
\end{align}
that has to be added to the RWA Hamiltonian $g_\ka$,
Eq.~(\ref{eq:g_period_2}), with $f' = 8F'/3\gamma C^3$.

If the rescaled field $f'$ is small, the term $\Delta g_\ka$ is small
compared to the depth of the wells of $g_\ka$. However, it can lead to
a significant change of the switching rates and, most importantly,
make the switching rates $\sigma_\ka\rightarrow -\sigma_\ka$ different
for $\sigma_\ka=1$ and $\sigma_\ka=-1$. In the stationary state, this
will lead to a difference of the well populations. In the classical
regime, where the interstate switching is thermally activated, the
change was discussed previously \cite{Ryvkine2006a}. We will show in
the following sections that, in the quantum regime, too, in several
cases of interest the major effect of the drive is to change the
quantum activation energy $R_{\sigma_\ka}$ compared to its value
$R_{\sigma_\ka}^{(0)} $ in the absence of the drive (similar to the
case of the switching rate, the notation $R_{\sigma_\ka}$ is used for
the value of $R$ for the oscillator $\ka$ in the state $\sigma_\ka$).
This leads to a corresponding change of the switching rate
$W_{\sigma_\ka}$,
\begin{align}
\label{eq:LS}
&W_{\sigma_\ka}\propto \exp[-R_{\sigma_\ka}/\lambda]\:, \qquad R_{\sigma_\ka} = R_{\sigma_\ka}^{(0)} + \Delta R_{\sigma_\ka}\:, \nonumber\\
 &\Delta R_{\sigma_\ka} = f'\sigma_\ka(\chi_{Q\ka}\sin \varphi_\ka  +\chi_{P\ka}\cos\varphi_\ka)\:.
\end{align}

In analogy to the classical case, we introduced the logarithmic
susceptibilities $\chi_{Q\ka}$ and $\chi_{P\ka}$ for the variables
$Q_\ka$ and $P_\ka$ of the mode $\ka$.  Note that to simplify the
further analysis we use a notation that differs from the one used in
Ref.~\onlinecite{Ryvkine2006a}.  To be specific, these
susceptibilities will be calculated for the well $\sigma_\ka=1$.  They
give the change of the logarithm of the switching rate
$W_{\sigma_\ka}$ linear in the drive $f'$.  The change of the rate can
be large even if $|\Delta R_{\sigma_\ka} |\ll R_{\sigma_\ka}^{(0)}$
provided $|\Delta R_{\sigma_\ka} |\gg \lambda$. By symmetry, the sign
of the change is opposite for the two different wells, and therefore
$\Delta R_{\sigma_\ka}\propto \sigma_\ka$.

In the previous discussion we tacitly implied that the
relaxation-induced broadening $\kappa\lambda$ of the eigenvalues of
$g_\ka$, i.e., of the scaled quasienergy levels, is small compared to
the level spacing. In this case, as we indicated, for isolated
oscillators these are the intrawell states close to the bottom of the
wells of $g_\ka$ that are primarily occupied. In the Wigner
representation, the probability distribution $\rho_\ka(Q_\ka,P_\ka)$
of an isolated oscillator has peaks, which are centered close to the
minima of $g_\ka(Q_\ka,P_\ka)$ and have a width $\sim \lambda^{1/2}$
\cite{Marthaler2006,Dykman2007}. However, if the level broadening is
not small, the Wigner distribution $\rho_\ka$ can still be
double-peaked. In the absence of a drive $F'$ it is symmetric. The
peaks are located at
$(\sigma_\ka Q_\ka^{(0)},\sigma_\ka P_\ka^{(0)})$, with
$\sigma_\ka=\pm 1$, and
\begin{align}
\label{eq:equilibrium}
&Q_\ka^{(0)}=(\mu_\ka-\mu_{B})^{1/2}\cos\Phi_\ka\:, \quad \mu_{B}=-(1-\kappa^2)^{1/2}\:,\nonumber\\
&P_\ka^{(0)} = (\mu_\ka-\mu_{B})^{1/2}\sin\Phi_\ka\:,\quad  \Phi_\ka = \arctan\frac{\kappa}{1-\mu_{B}}\:.
\end{align}
The coordinates $Q_\ka^{(0)}$, $P_\ka^{(0)}$ are given by the stable
stationary solution of the Langevin equation (\ref{eq:full_Langevin})
disregarding the coupling between the modes and the quantum noise.
The coefficient $\mu_{B}$ is the value of $\mu_\ka$ at the bifurcation
where the zero-amplitude state $Q_\ka=P_\ka=0$ becomes dynamically
unstable and the two stable period-2 states emerge. In the model we
use here, where the decay rate in the scaled time is the same for all
modes, $\mu_{B}$ is also the same for all modes.

\subsection{Switching rates for coupled modes}
\label{sec:Ising}

The separation of the time scales of relaxation and interwell
switching allows one to use the logarithmic susceptibilities to
describe the effect of a weak interaction between the oscillators. In
the absence of interaction, as seen from the master equation
(\ref{eq:master_full}), on a time scale long compared to
$\kappa^{-1}$, the dynamics of the modes can be described as rare
uncorrelated switching events between the wells. Most of the time each
oscillator spends in close vicinity of
$\pm(Q_\ka^{(0)},P_\ka^{(0)})$. We emphasize that, in each of these
states, the time-translation symmetry is broken.

The major effect of a weak interaction is that, if one oscillator is
in a given state $\sigma=\pm 1$, it lifts the time-translation
symmetry for the oscillators it is coupled to. In fact, it acts
exactly like a driving force $\propto F'$, as it also oscillates at
frequency $\omega_F/2$. Put differently, for any given oscillator
$\ka$, the oscillators $\ka'$ with $\ka'\neq \ka$
act as a drive at frequency $\omega_F/2$. The phase of this drive is
determined by the states $\sigma_{\ka'}$ of these oscillators. By
comparing the expressions for the change of $g_\ka$ due to an external
drive (\ref{eq:extra_field}) with the expression
(\ref{eq:coupling_RWA}) for the coupling term $g_c$, we see that the
switching rates between the states of the considered oscillator $\ka$
have the form
\begin{align}
\label{eq:coupled_rates}
&W_{\sigma_\ka}=W_{\sigma_\ka}^{(0)}\exp\left[-\sigma_\ka \sumprime{\ka'}J_{\ka \ka '} \sigma_{\ka'}/\lambda{}\right]\nonumber\\
&J_{\ka\ka'}=V_{\ka\ka'}[\chi_{Q\ka}Q_{\ka'}^{(0)} + \chi_{P\ka}P_{\ka'}^{(0)}].
\end{align}
Here we have approximated the dynamical variables $Q_{\ka'}, P_{\ka'}$
of the oscillators with $\ka'\neq\ka$ by their most probable values
$\sigma_{\ka'}Q_{\ka'}^{(0)}, \sigma_{\ka'}P_{\ka'}^{(0)}$.

Equation~(\ref{eq:coupled_rates}) maps the problem of the coupled
parametric oscillators onto a problem of coupled Ising spins. The
effect of the spin coupling is to modify the rates of switching
between the states of individual spins. If the oscillators slightly
differ in frequency or one of the other parameters, the spin-coupling
parameters are asymmetric, $J_{\ka\ka'}\neq J_{\ka'\ka}$. This is
because the equilibrium positions of different oscillators in the
rotating frame $(Q_\ka, P_\ka)$ are different and so are also the
logarithmic susceptibilities.

\subsection{The stationary distribution. Mapping onto the Ising model}
\label{subsec:Ising}

We will now look at the evolution of the distribution
$w(\sigma_1,\sigma_2,\ldots)$
of the states $\{\sigma_\ka \}$ of the system of effective spins.
The distribution changes due to
switching of the spins, that is, of the individual oscillators, with
the rates from Eq.~(\ref{eq:coupled_rates}). Given that the switching events
are independent, the function $w$ evolves according to the balance equation
\begin{align}
\label{eq:balance_populations}
  \dot w =& -\sum_\ka W_{\sigma_\ka}w (\sigma_1,\ldots,\sigma_\ka,\ldots) \nonumber\\
 &+\sum_\ka W_{-\sigma_\ka}w(\sigma_1,\ldots,-\sigma_\ka,\ldots)\:.
\end{align}

Importantly, if the oscillators are different, implying
$J_{\ka\ka'}\neq J_{\ka'\ka}$, the system lacks detailed
balance. Indeed, consider the probability of a pair of switching
events that bring the system from the state with given
$(\sigma_\ka,\sigma_{\ka'})$ to $(-\sigma_\ka, -\sigma_{\ka'})$.  It
is easy to see that this probability then depends on which of the
spins, $\ka$ or $\ka'$, switches first for
$J_{\ka\ka'}\neq J_{\ka'\ka}$. The violation of detailed balance is a
generic feature of systems far from thermal equilibrium, and the
system of driven oscillators considered here is in this category.

The dynamics of the system greatly simplifies in the case of identical
oscillators or in the vicinity of the bifurcation point, see
Secs.~\ref{subsec:weak_damping} and \ref{sec:near_bifurcation_point},
so that $J_{\ka\ka'}=J_{\ka'\ka}$.  In this case the stationary
solution of Eq.~(\ref{eq:balance_populations}) is
\begin{align}
\label{eq:Ising}
&w_{\rm st} = Z^{-1}\exp\left[-H(\{\sigma_\ka\})/\lambda\right]\nonumber\\
&H(\{\sigma_\ka\})=-\frac{1}{2}\sum_{\ka \neq \ka'} J_{\ka \ka'} \sigma_\ka \sigma_{\ka'}\:.
\end{align}
This exactly corresponds to the statistical distribution of an Ising
system at effective quantum temperature $\lambda$; the
normalization constant $Z$ plays the role of the partition
function. If the coupling constants of the oscillators
$\ep_{\ka\ka'}\propto V_{\ka \ka'} \propto J_{\ka \ka'} $ are
positive, the coupling is ``ferromagnetic'': in the most probable
state the values of $\sigma_\ka$ are the same for all oscillators,
that is, the oscillators vibrate in phase. This is intuitively clear:
if the oscillators attract each other, they will try to synchronize
into a state where they all vibrate in phase. Whether the system
reaches this fully ordered state is determined by the standard results
for the ferromagnetic Ising model.

In the opposite case where $\ep_{\ka\ka'}$ are negative,
Eq.~(\ref{eq:Ising}) maps the system of parametric oscillators onto an
antiferromagnetic Ising system. We emphasize that in the system of
oscillators that are currently studied, both the strength of the
coupling, and often its sign, can be independently controlled.

\section{Quantum logarithmic susceptibility}
\label{sec:LS}

To find the coupling parameters $J_{\ka\ka'}$ one has first to
calculate the logarithmic susceptibility of an {\it isolated oscillator}.
A general approach to such a calculation is based on
solving the variational problem for the exponent of the switching rate
$R_{\sigma_\ka}$, which can be formulated using the master equation
(\ref{eq:master_full}).  We will use simpler approaches, which apply
in the limiting cases. We will start with the case of weak damping,
where the broadening of the quasienergy levels is small compared to
the level spacing.

\subsection{Weak-damping limit}
\label{subsec:weak_damping}

In the weak-damping limit, we write the master equation
(\ref{eq:master_full}) for the isolated oscillator $\ka$ as a balance
equation for the populations $\rho_{m_\ka }$ of the eigenstates
$|m_\ka \rangle$ of the RWA Hamiltonian
\begin{align}
\label{eq:tilde_g}
G{}_{\ka}= g_{\ka}+\Delta g_{\ka}\:.
\end{align}
The operator $G_\ka$ describes an isolated parametric oscillator $\ka$
driven additionally by a weak field at half the frequency of the
parametric drive. The term $\Delta g_\ka$ is given by
Eq.~(\ref{eq:extra_field}); it is proportional to the weak field
amplitude $F'$.

For small $\lambda$ and $F'$, the function $G{}_\ka(Q_\ka,P_\ka)$ has
two slightly asymmetric wells (enumerated by $\sigma_\ka=\pm 1$),
cf. Fig.~\ref{fig:vibrations}. There are many eigenstates inside each
of the wells for $\lambda\ll 1$. We consider the states
$|m_\ka\rangle$ inside one of the wells and number them so that
$m_\ka =0$ corresponds to the lowest state.  Coupling to a thermal
reservoir leads to transitions
$|m_\ka+k_\ka\rangle \to |m_\ka\rangle$.  From
Eq.~(\ref{eq:master_full}), the rates $\Lambda_{m_\ka+k_\ka\,m_\ka}$ of
such intrawell transitions are given by the squared matrix elements of
the operators $a_\ka, a^\dagger_\ka$ on the corresponding
wave functions,
\begin{align}
\label{eq:transition_matrix_elements_general}
\Lambda_{m_\ka+k_\ka \, m_\ka} &=
2\kappa (\bar n +1)\left|\langle m_\ka|a_\ka|m_\ka + k_\ka\rangle\right|^2\nonumber\\
&+  2\kappa\bar n \left|\langle m_\ka +k_\ka|a_\ka|m_\ka\rangle\right|^2.
\end{align}
Generally, the rates of transitions with $k_\ka > 0$
are higher than with $k_\ka < 0$, i.e., the system is more likely to
move to eigenstates with lower $G_\ka$. This corresponds to the minima
of $G_\ka(Q_\ka,P_\ka)$ being stable states of the oscillator
in the classical limit.

However, even for zero temperature, in contrast to equilibrium
systems, transitions away from the minima of $G_\ka$, i.e., with
$k_\ka < 0$, have nonzero rates. They lead to the probability for an
oscillator, starting from deep inside of a well, to reach the
intrawell states near the top of the barrier of $G_\ka$. From there,
the oscillator will end up in each of the two wells with probability
$\sim 1/2$. The exponent of the switching rate $W_{\sigma_\ka}$ is
thus determined by the population of the intrawell states
$|m_\ka\rangle$ of the $\sigma_\ka$-well near the top of the
barrier. The logarithmic susceptibility describes the linear
dependence of the exponent of this population on the field $F'$.

It is convenient to seek the state populations $\rho_{m_\ka}$ in the
form $\rho_{m_\ka}=\exp[-R\bigl(G{}(m_\ka)\bigr)/\lambda]$, where
$G{}(m_\ka)$ is the eigenvalue of $G{}_\ka$ in the state
$|m_\ka\rangle$. This form of $\rho_{m_\ka}$ is reminiscent of a
Boltzmann distribution, with $\lambda$ playing the role of the
temperature and $R(G{})$ playing the role of the energy. In the
considered nonequilibrium case, $R$ is not a linear function of
$G{}$. The populations $\rho_{m_\ka}$ strongly vary with the level
number $m_\ka$. However, generally, the function $R(G{})$ is smooth even
for small $\lambda$.

The equation for $R(G{})$ can be obtained from the master equation in
the eikonal (WKB) approximation.
It is seen from Eq.~(\ref{eq:transition_matrix_elements_general}) that
the intrawell transition rates $\Lambda_{m_\ka+k_\ka\,m_\ka}$ fall off
exponentially fast with $|k_\ka|$. One can then expand
$R\bigl(G{}(m_\ka+k_\ka)\bigr)\approx R\bigl(G{}(m_\ka)\bigr)+\lambda
k_\ka [\omega(G{})dR/dG{}]_{m_\ka}$, which is essentially the 
eikonal approximation. Here, $\omega(G{})$ is the
frequency of classical vibrations with a given $G{}$; we used that
$\omega(G{}_{m_\ka}) =[G{}(m_\ka + 1)-G{}(m_\ka)]/\lambda +
o(\lambda)$; $[\cdot]_{m_\ka}$ indicates that the function of $G{}$ is
evaluated for $G{}=G{}(m_\ka)$.

In deriving the equation for $R(G{})$ one should keep in mind that
changing $m_\ka$ by $k_\ka$ in $\Lambda_{m_\ka+k_\ka\,m_\ka}$ leads to
a small change that can be disregarded for $m_\ka \gg 1$ and
$|k_\ka|\ll m_\ka$.  Yet another fact is that the intrawell
distribution $\rho_{m_\ka}$ is formed on a timescale
$\sim 1/\kappa \ll 1/W_{\sigma_\ka}$. Thus, for times
$\ll 1/W_{\sigma_\ka}$ the populations of the intrawell states are
given by the stationary solution of the balance equation. Using these
arguments, one can derive from Eq.~(\ref{eq:master_full}) the balance
equation for the intrawell state populations as
\begin{align}
\label{eq:balance}
  \sum_{k_\ka} \Lambda_{m_\ka + k_\ka\,m_\ka}&\left\{1-
  \exp\left[-k_\ka\bigl[\omega(G{})dR/dG{}\bigr]_{m_\ka}\right]\right\}
   =0\:.
\end{align}
In the WKB approximation that we used, the matrix elements of the
lowering operator $a_{\ka}$ in
Eq.~(\ref{eq:transition_matrix_elements_general}) can be written as
\begin{align}
\label{eq:WKB_Fourier_components}
&a_{k_\ka}(m_\ka)\equiv \langle k_\ka+m_\ka|a_\ka|m_\ka\rangle \nonumber \\
&
=\frac{1}{2\pi}\int_0^{2\pi}d\phi \exp(-ik_\ka\phi)a_\ka\bigl(G{}(m_\ka)\big\rvert \phi\bigr)\:,
\end{align}
where $a_\ka(G{}\big\rvert\phi)$ is the value of
$a_\ka=(2\lambda)^{-1/2}(Q_\ka+iP_\ka)$
calculated as a classical function of the phase $\phi=\omega(G{})\tau$ on the
classical intrawell trajectory with a given $G{}$.

Equation~(\ref{eq:balance}) is an algebraic equation for
$\pi_\ka = \omega(G{}_\ka) dR/dG{}_\ka$.  In the absence of an extra
drive $\propto F'$, it was derived and solved in
Ref.~\cite{Marthaler2006}. Importantly, in this equation one can treat
$\lambda m_\ka \equiv I_\ka$ as a continuous variable. To leading
order in $\lambda$, $\omega(G{}_\ka)= dG{}_\ka/dI_\ka$. The variable
$I_\ka= (2\pi)^{-1}\oint P_\ka dQ_\ka$ is the classical action for the
intrawell orbit with a given $G_\ka$. Equation~(\ref{eq:balance}) does
not contain the effective Planck constant $\lambda$; it gives
$\pi_\ka =dR/dI_\ka$ as a function of the continuous variable $I_\ka$.

The change of $R$ due to the perturbation $\Delta g_{\ka}$ can be
found by finding the change $\Delta \Lambda_{m_\ka+k_\ka\,m_\ka}$ of
the intrawell transition rates compared to their values
$\Lambda^{(0)}_{m_\ka+k_\ka\,m_\ka}$ for $\Delta g_\ka =0$.  In turn,
the rate change comes from the change of the matrix elements
$a_{k_\ka}(m_\ka)$. The correction to $a_{k_\ka}(m_\ka)$ of the
first-order in $\Delta g_\ka$ can be obtained from
Eq.~(\ref{eq:WKB_Fourier_components}) using the classical equations of
motion for $Q_\ka, P_\ka$ with the perturbed effective Hamiltonian
$g_\ka(Q_\ka,P_\ka)+\Delta g_\ka(Q_\ka,P_\ka)$, which is a standard
problem of classical nonlinear mechanics \cite{Arnold1989}. An
important simplification is that, in the limit of weak damping, the
value of the momentum in a stable state is $P_\ka^{(0)}=0$. Therefore
the coupling parameters $J_{\ka\ka'}$ in Eq.~(\ref{eq:coupled_rates})
are determined only by the $\chi_{Q_\ka}$-component of the logarithmic
susceptibility. As a consequence, as seen from
Eq.~(\ref{eq:extra_field}), when calculating the correction to
$a_{k_\ka}(m_\ka)$ we can limit the analysis to
$\Delta g_\ka = -f'Q_\ka$, i.e., $\varphi_\ka = \pi/2$ in
Eqs.~(\ref{eq:extra_field}) and (\ref{eq:LS}).

Since the leading-order corrections to the intrawell transition rates
are linear in $f'\propto F'$, so is also the leading-order correction
$\Delta\pi_\ka(I_\ka)$ to the unperturbed value
$\pi_\ka^{(0)}(I_\ka)$. From Eq.~(\ref{eq:balance}) it has the form
\begin{align*}
&\Delta\pi_\ka(I_\ka)= -\sum_{k_\ka} \Delta \Lambda_{m_\ka + k_\ka\,m_\ka}\left\{1-\exp[-k_\ka \pi_\ka^{(0)}(I_\ka) ]\right\}\nonumber\\
&\times \left\{\sum_{k_\ka} k_\ka \Lambda^{(0)}_{m_\ka + k_\ka\, m_\ka}\exp[-k_\ka\pi_\ka^{(0)}]\right\}^{-1}\,,
\end{align*}
where $ m_\ka= I_\ka/\lambda$ and the rates
$\Lambda^{(0)}, \Delta\Lambda$ are considered continuous functions of
$I_\ka$.

The logarithmic susceptibility is  
\begin{align}
\label{eq:LS_underdamped}
\chi_{Q\ka} = \frac{1}{f^\prime}\int_0^{I_{\ka\,\max}}\Delta\pi_\ka(I_\ka)dI_\ka\:.
\end{align}
As indicated above, it is assumed here that $\Delta\pi_\ka$ is
calculated for the $\sigma_\ka=1$-well of the oscillator (the well of
$g_\ka(Q_\ka,P_\ka)$ with the minimum at $Q_\ka^{(0)}>0$). The upper
limit $I_{\ka\,\max}$ is the value of the mechanical action in this
well at the barrier top of $g_\ka$. In the case of weak damping, the
logarithmic susceptibility depends on two parameters, $\mu_\ka$ and
$\bar n$.  Generally, Eqs.~(\ref{eq:LS_underdamped}) and
(\ref{eq:coupled_rates}) suggest that $J_{\ka\ka'}$ is not symmetric
with respect to the interchange $\ka\leftrightarrow \ka'$ in the
presence of disorder in the oscillator system.

In the absence of the drive $\propto F'$, the
assumption of $R$ being a smooth function of $g_\ka$ breaks down for a
certain range of $\mu_\ka$ in a very narrow range of temperatures; for
a resonantly driven oscillator this range was found to be limited to
$\exp(-1/\lambda)\ll \bar n \ll \lambda^{3/2}$ \cite{Guo2013}. It
is important that, for $\bar n\to 0$, the perturbation $\Delta g_\ka$
does not break the smoothness of $R(g_\ka)$. One can see this by
showing that the exponent of the decay of the intrawell transition
rates $\Lambda_{m_\ka + k_\ka\,m_\ka}$ with $|k_\ka|$ is weakly
modified by a weak perturbation. The analysis of the decay is somewhat
involved and will be presented elsewhere. Here we only note that a
weak change of the decay exponent of the rates means that the sum over
$k_\ka$ in Eq.~(\ref{eq:balance}) remains converging rapidly for
$R(G{}_\ka)$ close to its value in the absence of the perturbation.

\subsubsection{Approaching the bifurcation point}
\label{subsec:bifurcation_underdamped}

The analysis of Eq.~(\ref{eq:balance}) is greatly simplified if
$dR/dI_\ka \ll 1$. This happens for $\mu_\ka$ close to the
bifurcation point $\mu_{B}$, see Eq.~(\ref{eq:equilibrium}); in the
limit of weak damping, $\mu_{B}\to -1$. In this case one can expand
the exponential factor in Eq.~(\ref{eq:balance}) to second order
in $dR/dI_\ka\equiv \omega(G_\ka)dR/dG{}_\ka$. In the absence of an
extra drive the calculation was described in
\cite{Marthaler2006}. However, it can be immediately generalized to
the case where such a drive is present, as in
Eq.~(\ref{eq:balance}). One then finds from Eqs.~(\ref{eq:balance})
and (\ref{eq:WKB_Fourier_components}) that, even before the
linearization with respect to $f'$, the resulting expression for
$dR/dI_\ka$ is similar to that for a classical oscillator,
\begin{align}
\label{eq:small_R_derivative}
&\frac{dR}{dI_\ka} =\frac{2\omega(G{}_\ka)}{2\bar n+1}\frac{2\pi I_\ka}{N(G{}_\ka)},
\quad I_\ka = \frac{1}{2\pi}\iint dQ_\ka dP_\ka\nonumber\\
&N(G{}_\ka) = \iint dQ_\ka dP_\ka[2(Q_\ka^2 + P_\ka^2)-\mu_\ka]\:.
\end{align}
The integration in the expressions for $N(G{}_\ka)$ and $I_\ka$ is
done over the interior of the contour $G_\ka (Q_\ka,P_\ka)=G{}_\ka$.
Equation (\ref{eq:small_R_derivative}) applies near the bifurcation
point because the frequency $\omega(G{}_\ka)$ is small,
$\omega(G{}_\ka)\leq 2\sqrt{\mu_{\ka}+1}\ll1$, and therefore
$dR/dI_\ka\ll 1$.  We note also that, in the expression for
$N(G_\ka)$, $Q_\ka^2$ and $P_\ka^2$ are small, which allows one to
easily find the ratio $I_\ka/N(G_\ka)$.

From Eq.~(\ref{eq:small_R_derivative}) 
\begin{align}
\label{eq:LS_underdamped_bifurcation}
&\chi_{Q\ka} =2(\mu_\ka +1)^{1/2}/(2\bar n +1), \qquad \mu_\ka +1\ll 1,\nonumber\\
&J_{\ka\ka'} = [2/(2\bar n +1)]V_{\ka\ka'}Q_\ka^{(0)}Q_{\ka'}^{(0)}\:.
\end{align}
The expression (\ref{eq:LS_underdamped_bifurcation}) for the
``spin-coupling'' parameters $J_{\ka\ka'}$ is bilinear in the
positions of the wells of the coupled oscillators
$Q_\ka^{(0)}=(\mu_\ka -\mu_B)^{1/2}$. It is symmetric even in the
presence of a weak disorder in the oscillator eigenfrequencies,
$J_{\ka\ka'}=J_{\ka'\ka}$. Therefore, near the bifurcation point, but
still in the range where the damping is small, the system of coupled
parametric oscillators maps onto the equilibrium Ising model.

\subsubsection{Classical limit}

The condition of applicability of Eq.~(\ref{eq:small_R_derivative}),
$dR/dI_\ka \ll 1$, applies also far from the bifurcation point if the
temperature is high, $\bar n\gg 1$. The logarithmic susceptibility as
a function of the only parameter $\mu_\ka$ in this case was found in
Ref.~\cite{Ryvkine2006a}. It is important that it is not proportional
to $Q_{\ka}^{(0)}$. Therefore in the classical limit
$J_{\ka\ka'}\neq J_{\ka'\ka}$ and the system of coupled parametric
oscillators does not map on the Ising model in the presence of disorder.

\section{Vicinity of the bifurcation point}
\label{sec:near_bifurcation_point}

The role of dissipation becomes increasingly more important as an
isolated damped oscillator approaches the bifurcation point, which in
the presence of dissipation is located at
$\mu_{B}=-(1-\kappa^2)^{1/2}$, cf. Eq.~(\ref{eq:equilibrium}).  For
$\mu_\ka -\mu_{B} \ll \kappa$, the logarithmic susceptibility in the
classical limit was calculated in Ref.~\cite{Ryvkine2006a}. It was
also shown \cite{Dykman2007} that the quantum dynamics near the
bifurcation point is similar to the classical dynamics. Therefore one
can show that the expression for the quantum logarithmic
susceptibility is similar to that for the classical one, which allows
finding the parameters $J_{\ka\ka'}$ for weakly coupled oscillators.

A better insight can be gained by formulating the problem of coupled
oscillators somewhat differently. We will assume that all oscillators
are close to the bifurcation point, i.e., that the condition
$\mu_\ka - \mu_{B}\ll \kappa$ holds for all $\ka$. It is then
convenient to rotate the variables by changing to the new coordinates
and momenta ${\tilde Q}_\ka$, ${\tilde P}_\ka $.
They are defined by
${\tilde Q}_\ka +i{\tilde P}_\ka = (Q_\ka +iP_\ka )\exp(-i\beta)$ with
$\beta=(\pi-\arcsin\kappa)/2$. In these variables,
\begin{align}
\label{eq:rotated_g}
g_{\ka} =& \frac{1}{4}(\tilde Q_\ka ^2 + \tilde P_\ka ^2 -\mu)^2 + \frac{1}{2}(\tilde P_\ka ^2 -\tilde Q_\ka ^2)\cos 2\beta \nonumber\\
&+\tilde P_\ka \tilde Q_\ka \sin 2\beta - \mu^2/4\:,
\end{align}
and $g_c = -\frac{1}{2}\sum_{\ka \neq \ka'} V_{\ka \ka'} (\tilde Q_\ka \tilde Q_{\ka'} + \tilde P_\ka \tilde P_{\ka'})$.

Rewriting Eq.~(\ref{eq:full_Langevin}) in the new variables, one
immediately finds that, over a dimensionless time
$ (2\kappa)^{-1}$, the variable ${\tilde P}_\ka $ relaxes to its
quasiequilibrium value
${\tilde P}_\ka \approx (\mu_B/\kappa){\tilde Q}_\ka
-\sumprime{\ka'}V_{\ka \ka'} \tilde Q_{\ka'}/2\kappa$, whereas the
relaxation of ${\tilde Q}_\ka $ is much slower. Such a separation of
time scales is characteristic of the dynamics near a bifurcation point
\cite{Guckenheimer1987}. The variable $\tilde Q_\ka$ is the analog of
a ``soft'' mode. Its fluctuations are much stronger than the fluctuations
of $\tilde P_\ka$. In other words, if one writes down the master
equation in the Wigner representation, the distribution over
${\tilde P}_\ka $ is much narrower than over ${\tilde Q}_\ka $, see
\cite{ Dykman2007}. One can disregard the fluctuations of
${\tilde P}_\ka $, and then the problem is reduced to the dynamics 
of one variable per oscillator. To leading order in $\mu_\ka-\mu_B$ the Langevin
equations (\ref{eq:full_Langevin}) take the particularly simple and
intuitive form
\begin{align}
\label{eq:adiabatic_parametric}
&\frac{d}{d\tau} {\tilde Q}_\ka  \approx -\frac{\partial \mathbb{U}(\{\tilde Q_\ka\})}{\partial \tilde Q_\ka} +{\tilde f}_\ka  (\tau), \nonumber\\
&\mathbb{U}(\{\tilde Q_\ka\}) = \frac{|\mu_B|}{\kappa}\sum_\ka\left[-\frac{1}{2}(\mu_\ka -\mu_B)\tilde Q_\ka^2  +\frac{1}{4\kappa^2}{\tilde Q}_\ka ^4\right] \nonumber\\
&-\frac{|\mu_B|}{2\kappa}\sum_{\ka\neq\ka'}V_{\ka \ka'} \tilde Q_\ka \tilde Q_{\ka'}\:,
\end{align}
where 
$ \tilde f_\ka  = f_{Q_\ka} \cos\beta +  f_{P_\ka}\sin\beta$ is a
$\delta$-correlated noise with
$\langle \tilde f_\ka (\tau) \tilde f_{\ka'}(\tau')\rangle = 2D{}\delta(\tau-\tau')\delta_{\ka \ka'}$, cf. Eq.~(\ref{eq:noise_correlators}). Importantly, $\tilde f_\ka$ 
can be considered to be a $c$-number, because there is only one
variable component of the noise for each oscillator. Moreover, since the
dynamics of each oscillator is described by only one variable, this
dynamics is {\it classical}. The only trace of the quantum formulation
is that the noise intensity $D$ is proportional to $\hbar$ for small  $\bar n$.

Equation (\ref{eq:adiabatic_parametric}) shows that, near the
bifurcation point, the dynamics of coupled quantum parametric
oscillators in the rotating frame maps onto the dynamics of a system
of coupled overdamped Brownian particles. Each particle moves in a
quartic bistable potential, and the coupling between the particles is
bilinear in their coordinates.

\subsection{The weak-coupling limit}
The behavior of the system (\ref{eq:adiabatic_parametric}) strongly
depends on the relation between two small parameters: the
coupling strength $|V_{\ka\ka'}|$ and the distance to the bifurcation
point $\mu_\ka - \mu_B$.  The results are particularly simple if
$|V_{\ka\ka'}|\ll \mu_\ka - \mu_B$ for all $\ka$. Here, in the absence
of coupling to other oscillators, the stable states of a $\ka$th oscillator
are $\sigma_\ka  \tilde Q_\ka^{(0)}$, 
\begin{align}
\label{eq:amplitude_bifurcation}
\tilde Q_\ka^{(0)}= \kappa(\mu_\ka -\mu_B)^{1/2}\:.
\end{align}
The noise $\tilde f_\ka$ leads to switching between the states
$\sigma_\ka = \pm 1$. The coupling to other oscillators modifies the
switching rate $W_{\sigma_\ka}$. As discussed earlier, for weak noise
intensity switching events are rare and, most likely, when one
oscillator switches, the oscillators it is coupled to are close to
their equilibrium positions (\ref{eq:amplitude_bifurcation}). Then,
the switching rate is given by the Kramers expression
\cite{Kramers1940} for a thermally activated transition over a
potential barrier, except that in the case considered here the origin
of the fluctuations is quantum \cite{Dykman2007}. To lowest order in
$V_{\ka\ka'}$,
\begin{align}
\label{eq:bif_switch_rate}
&W_{\sigma_\ka} = C_\ka\exp[-R_{\sigma_\ka}/\lambda{}], \qquad R_{\sigma_\ka} =  
R_{\sigma_\ka}^{(0)} + \Delta R_{\sigma_\ka}
\nonumber\\
 &R_{\sigma_\ka}^{(0)} = \frac{|\mu_B|(\mu_\ka - \mu_B)^2}{2(2\bar n +1)}, \quad \Delta R_{\sigma_\ka} = 
\sigma_\ka \sumprime{\ka'}J_{\ka \ka'} \sigma_{\ka'}, \nonumber\\
&J_{\ka \ka'}  = 2V_{\ka \ka'} |\mu_B \tilde Q_\ka^{(0)} \tilde Q_{\ka'}^{(0)}| /\kappa^2(2\bar n +1).
\end{align}
The prefactor in the switching rate in dimensionless time is
$C_\ka=|\mu_B|(\mu_\ka -\mu_B)/(\sqrt{2}\,\pi \kappa)$. We note that, surprisingly, Eq.~(\ref{eq:bif_switch_rate}) for $J_{\ka\ka'}$ goes over into Eq.~(\ref{eq:LS_underdamped_bifurcation}) for $J_{\ka\ka'}$ in the limit $\kappa\to 0$.

Equation (\ref{eq:bif_switch_rate}) can be obtained also
using the logarithmic susceptibility near the bifurcation point. We
note that $J_{\ka\ka'} = J_{\ka'\ka}$, and therefore the
stationary distribution of the system of coupled parametric
oscillators maps on the Ising model. Importantly, the correction
$\Delta R_{\sigma_\ka}$ falls off slower than $R_{\sigma_\ka}^{(0)}$ as the
oscillator approaches the bifurcation point and $\mu_\ka - \mu_B$
decreases. This means that the role of the coupling increases closer to the
bifurcation point.

\subsection{Stronger coupling: a ``time glass''}

Sufficiently close to the bifurcation point the condition of a very
weak coupling $|V_{\ka \ka'} |\ll \mu_\ka - \mu_B$ breaks down for
many, if not for all oscillators $\ka$. If this happens, i.e., if the
coupling is stronger, but still $|V_{\ka \ka'} |\ll |\mu_B|$ [as seen
from Eq.~(\ref{eq:equilibrium}), $|\mu_B|<1]$, the dynamics of the
coupled oscillators near the bifurcation point is described by
Eq.~(\ref{eq:adiabatic_parametric}), except that now this equation
cannot be solved by perturbation theory in $V_{\ka \ka'} $.

In the absence of noise, Eq.~(\ref{eq:adiabatic_parametric}) has stable
stationary solutions, which are inversion-symmetric
($\tilde Q_\ka \to -\tilde Q_\ka$), as expected, and correspond to the
broken time-translation symmetry. However, these are no longer weakly
perturbed single-oscillator states
(\ref{eq:amplitude_bifurcation}). Rather, these states are formed as a
result of the coupling.
They are located at the minima of the potential ``landscape''
$\mathbb{U}(\{\tilde Q_\ka\})$. Generally, this landscape has multiple
minima with depth $\sim \kappa |\mu_B |V_{\ka\ka'}^2$, as seen from
Eq.~(\ref{eq:adiabatic_parametric}). If this depth largely exceeds the
noise intensity $D=\lambda\kappa(2\bar n +1)/2$, once the system is
near a minimum, it will stay there for a long time.
This would mean that we can have various types of {\it many-body}
metastable broken-symmetry states, a spin-glass analog in the time
domain.

\section{Quantum phase transition in the lattice of parametric oscillators}
\label{sec:QPT}

\subsection{Many-body ``ground" state}

We now consider a closed system of quantum parametric oscillators,
i.e., we assume that the oscillators are isolated from a thermal
reservoir.  For a single quantum oscillator, the possibility to have a
broken-symmetry state is a consequence of the exact degeneracy of the
eigenvalues of $g_\ka$ for a discrete set of the values of the ratio
$\mu_\ka/\lambda$ \cite{Marthaler2007a,*Zhang2017a}. A combination of
the corresponding eigenstates is a period-2 state.

For a system of coupled oscillators the situation is different. We
will consider the simplest case where the oscillators are identical,
form a periodic lattice, and the coupling is ferromagnetic. To allow
for two- or three-dimensional systems, we will
index the oscillators by a vector $\varkab$, which can be thought of
as the position of the corresponding oscillator. Our primary interest
will be the spectrum of excitations in the system and how it evolves
on varying the control parameter $\mu$, which is now the same for
all oscillators, $\mu_\varkab = \mu$.

The extrema of the RWA Hamiltonian (\ref{eq:scaled_Hamiltonian})
${\mathbb G}({Q_\varkab,P_\varkab}\}) $ of the coupled oscillators are
given by the equation
\begin{align}
\label{eq:extremum_g2}
&Q_\varkab (Q_\varkab ^2 + P_\varkab ^2 -\mu  -1)- \sumprime{\varkab'}V_{\varkab \varkab'} Q_{\varkab'}=0\:,\nonumber\\
&P_\varkab (Q_\varkab ^2 + P_\varkab ^2 -\mu  +1)- \sumprime{\varkab'}V_{\varkab \varkab'} P_{\varkab'}=0\:.
\end{align}
For a strongly detuned or weak driving field, $-\mu \gg 1$, the oscillators can be prepared in their ground state. This corresponds to  the solution of the above equation
\begin{align}
\label{eq:far_from_QPT}
Q_\varkab^{(0)} = P_\varkab^{(0)}=0, \qquad {\mathbb G}^{(0)}=0  \quad (\mu<\mu_{\rm QPT})\:,
\end{align}
where $\mu_{\rm QPT}$ is defined below in Eq.~(\ref{eq:critical_mu});
we disregard quantum corrections to ${\mathbb G}^{(0)}$. Excitations
in this regime can be obtained by linearizing the equations of motion
$\dot Q_\varkab = \partial {\mathbb G}/\partial P_\varkab$,
$\dot P_\varkab = -\partial {\mathbb G}/\partial Q_\varkab$ about
$Q_\varkab^{(0)} = P_\varkab^{(0)}=0$ and seeking the solution for the
increments of $Q_\varkab$, $P_\varkab$ in the standard form
$\delta Q_\varkab = \delta Q(\kb)\exp(i\kb\varkab)$,
$\delta P_\varkab = \delta P(\kb)\exp(i\kb\varkab)$. They are ``optical
phonons'' with frequencies
\begin{align}
\label{eq:phonons_above_QPT}
\omega^{(0)}(\kb)& = \left\{ [\mu+V(\kb)]^2-1\right\}^{1/2},\nonumber\\
V(\kb)& = \sumprime{\varkab'}V_{\varkab \varkab'}\exp[i\kb (\varkab'-\varkab)]\:.
\end{align}
The Fourier components of the coupling parameters have the property
$V(\kb) = V^*(\kb)$: this is because
$V_{\varkab\varkab'}=V_{\varkab'\varkab}$ and $V_{\varkab\varkab'}$ is
translationally invariant. Thus, for sufficiently large $-\mu$, the
frequencies (\ref{eq:phonons_above_QPT}) are real. They correspond to
the (scaled) frequencies of the undriven coupled oscillators with the
Hamiltonian $H_0+H_c$, Eqs.~(\ref{eq:H0}) and (\ref{eq:coupling}), shifted by
$-\omega_F/2$. We note that there is only one branch of phonons in the
system of coupled oscillators even in the absence of the periodic
drive, as each oscillator has only one degree of freedom.

The spectrum (\ref{eq:phonons_above_QPT}) is gapped, as illustrated in
the left panel in Fig.~\ref{fig:QPT}. For small $k$,
\begin{align}
\label{eq:gap_above_QPT}
&\omega^{(0)}(\kb)\approx  \omega^{(0)}(0) - \frac{\mu+V(0)}{2\omega^{(0)}}\sumprime{\varkab'}V_{\varkab\varkab'}[\kb(\varkab-\varkab')]^2\:,\nonumber\\
&\omega^{(0)}(0) = [(2+\mu_{\rm QPT}-\mu)((\mu_{\rm QPT}-\mu)]^{1/2} \quad (\mu<\mu_{\rm QPT})\:,
\end{align}
where
\begin{align}
\label{eq:critical_mu}
\mu_{\rm QPT}= -1 -V({\bf 0})
\end{align}
(we note that  $\mu_{\rm QPT} < - 1$).

As $\mu$ increases and approaches $\mu_{\rm QPT}$, the spectral gap
$\omega^{(0)}(0)$ decreases. For $\mu=\mu_{\rm QPT}$ the gap goes to
zero and the spectrum of the Floquet phonons becomes linear for
$k\to 0$, see the central panel in Fig.~\ref{fig:QPT}:
$\omega^{(0)}(\kb)\to \omega_{\rm QPT}(\kb)$. For small $k$
\begin{align}
\label{eq:critical_spectrum}
\omega_{\rm QPT}(\kb) \approx  \left\{\sumprime{\varkab'}V_{\varkab\varkab'}[\kb(\varkab-\varkab')]^2\right\}^{1/2}\propto k.
\end{align} 

For $\mu>\mu_{\rm QPT}$ the extremum (\ref{eq:far_from_QPT}) is no
longer the minimum of the RWA Hamiltonian ${\mathbb G}$. As seen from
Eq.~(\ref{eq:extremum_g2}), ${\mathbb G}$ has two equally deep minima
of depth $ {\mathbb G}^{(0)}$, which are located at
\begin{align}
\label{eq:symmetric_solution}
&Q_\varkab  = \pm  Q^{(0)}, \; P_\varkab =0; \qquad Q^{(0)} = (\mu-\mu_{\rm QPT})^{1/2}\:, \nonumber\\
&{\mathbb G}^{(0)}= -(\mu-\mu_{\rm QPT})^2/4 \quad (\mu>\mu_{\rm QPT})\:.
\end{align}
We checked numerically for short chains with nearest-neighbor coupling
that Eq.~(\ref{eq:symmetric_solution}) provides the global minimum of
${\mathbb G}$.

The solution (\ref{eq:symmetric_solution}) describes two degenerate
quantum-coherent period-2 states of the system of coupled
oscillators. Excitations about these states can be found by
linearizing the quantum equations of motion for $Q_\varkab$ and
$P_\varkab$, as it was done above for excitations about the state
(\ref{eq:far_from_QPT}). The frequencies of the corresponding Floquet
phonons are
\begin{align}
\label{eq:mode_frequencies}
&\omega^{(0)}(\kb)= [1-\mu_{\rm QPT} -V(\kb)]^{1/2}\nonumber\\
&\times [2(\mu-\mu_{\rm QPT} )+V({\bf 0}) - V(\kb)]^{1/2} \quad (\mu>\mu_{\rm QPT}).
\end{align}
The spectrum (\ref{eq:mode_frequencies}) is gapped, cf. the right
panel in Fig.~\ref{fig:QPT}, with
$\omega^{(0)}({\bf 0}) = 2(\mu-\mu_{\rm QPT})^{1/2}$; the difference
$\omega^{(0)}(\kb)-\omega^{(0)}({\bf 0})$ is quadratic in $k$ for
small $k$, as in the case $\mu < \mu_{\rm QPT}$.

The evolution of the system as $\mu$ increases from below to above
$\mu_{\rm QPT}$ corresponds to a quantum phase transition to a
many-body period-2 state. If this evolution occurs as $\mu$
is slowly increased in time, the transition should have the familiar
features associated with the creation of topological defects due to
the nonadiabaticity that occurs where the excitation gap approaches
zero, see \cite{Dziarmaga2010,Polkovnikov2011}. Still the resulting
state of the system has a broken time-translation symmetry.

A transition through the critical point can be performed by changing
the frequency or the amplitude of the driving force (or both). The
parameter scaling used above was done for a nonzero field amplitude. An
alternative scaling that allows turning the field on from zero is
described in Appendix \ref{sec:scaled_amplitude}.

\section{Conclusions}
\label{sec:conclusions}

The results of this paper show that the Floquet dynamics of coupled
quantum oscillators can display breaking of the discrete
time-translation symmetry imposed by a periodic field. This symmetry
breaking occurs when the frequency of the driving field is close to
twice the eigenfrequencies of the oscillators. The broken-symmetry
state corresponds to the phases of the parametrically excited
vibrations of different oscillators taking correlated values; the
system has an equivalent state where all these phases differ by $\pi$.
This can be contrasted with the case of uncoupled oscillators, where
the vibration phases are uncorrelated and on average the symmetry in a
large system is not broken.

The symmetry breaking does not require disorder in the system.
Because the energy spectrum consists of narrow slightly nonequidistant bands, weak driving
does not lead to heating of the system even
in the absence of dissipation.  Transitions between the degenerate
broken-symmetry states correspond to a phase slip. For a many-body
state, collective phase slips are rare and the lifetime of the
broken-symmetry state is extremely long.

In contrast to a single quantum-coherent (non-dissipative) parametric
oscillator, where the symmetry breaking is possible but requires fine
tuning of the interrelation between the amplitude and frequency of the
driving field, for coherent coupled oscillators no fine-tuning is
needed.  The symmetry-breaking transition in this case is a quantum
phase transition and occurs as the amplitude or frequency of the
driving field are changed so that they cross the corresponding
critical values.

In the presence of dissipation, an individual oscillator $\ka$ has two
metastable broken-symmetry states with opposite phases, and quantum
fluctuations lead to transitions between these states.  The coupling
modifies the rates of these transitions. In the considered case of
weak coupling between the oscillators, the rates could be found using
the logarithmic susceptibility of an isolated oscillator that
describes its response to a weak extra field. The coupled oscillators
map on a system of coupled spins $\{\sigma_\ka\}$. The different
broken-symmetry states of an oscillator $\ka$ correspond to different
values $\sigma_\ka=\pm 1$. For a large system and if the coupling is
not too weak, a stationary state is formed where the phases of all
oscillators (the values of $\sigma_\ka$ with different $\ka$) are
strongly correlated, if the effective dimension of the system is
larger than one. This is a broken-symmetry state.

If the driving field parameters are sufficiently close to the
bifurcation point, the coupled spins can be effectively described by
an Ising model with effective temperature $\propto \hbar$, for low
temperature. The mapping applies if the oscillators are slightly
different, i.e., if the system is disordered.  It holds both if the
oscillators are underdamped and if they are closer to the bifurcation
point, so that the damping becomes important.

The mapping onto the Ising model breaks down in two important cases. One
case is where the system is disordered and is far from the bifurcation
point. Here, one can still map coupled parametric oscillators onto
coupled spins, but the spin dynamics lacks detailed balance. In the
stationary state, there is a microscopic current in the ``spin
space''. This is a consequence of the oscillators being far from
thermal equilibrium. To the best of our knowledge, the dynamics of
Ising spins in the absence of detailed balance has not been explored,
and coupled parametric oscillators provide a platform for studying
this dynamics.

The other case is where the oscillators are close to the bifurcation
point but their coupling may no longer be assumed weak. Such a regime
invariably emerges as the bifurcation point is approached: There, each
oscillator becomes more and more sensitive to perturbations, including
coupling to other oscillators. In this regime the dynamics can be
mapped onto that of coupled overdamped Brownian particles driven by
quantum noise. For low temperatures the noise intensity is
$\propto \hbar$. The resulting ``potential landscape" has multiple
metastable minima. Each of them corresponds to a broken time symmetry
state of the system.

The rich pattern of symmetry-broken states described here and the
possibility of controlling them by varying the parameters of the
driving field makes the system of parametric quantum oscillators
attractive for studying quantum ``time-crystal'' phenomena.  As
mentioned in the introduction, an appropriate platform for such
studies is provided, for example, by various 
well-characterized mesoscopic oscillatory systems with controlled coupling between the modes.
The results bear not only on the time symmetry breaking, but also on the general problems of quantum physics far from thermal equilibrium, including such important problems as
nonequilibrium quantum phase transitions, quantum-fluctuations induced
microscopic currents in the stationary state, and quantum diffusion in
a potential landscape.


\acknowledgments
MID acknowledges the warm hospitality at the
University of Konstanz and the partial support from the Department of
Physics and the Zukunftskolleg Senior Fellowship. His research was
also supported in part by the National Science Foundation (Grant
No. CMMI-1661618). CB and NL acknowledge financial support by the
Swiss SNF and the NCCR Quantum Science and Technology. Y.Z. was supported by the National Science Foundation (DMR-1609326).

\appendix
\section{Turning up the driving amplitude}
\label{sec:scaled_amplitude}

To develop a formulation that will allow us to see how the quantum
phase transition occurs on increasing the amplitude of the driving
force, we introduce a scaling amplitude $F_s$.  The dimensionless
parameters of the dynamics are
\begin{align}
\label{eq:period2with_F}
&\fp = \frac{F}{F_s}\:, \quad \mu_\ka' = \frac{\omega_{F}(\omega_{F}-2\omega_\ka)}{F_s}\sgn\gamma\:, \nonumber\\
&C' = \vert 2F_s/3\gamma\vert^{1/2}\:,\quad\lambda'=3|\gamma|\hbar/\omega_FF_s\:,
\end{align}
and we define the slow variables as $U^\dagger(t)[q_\ka
+(2i/\omega_{F})p_\ka ]U(t)=-iC'(Q_\ka +iP_\ka )e^{-i\omega_Ft/2}$. This leads to $U^\dagger \,H \,U - i\hbar U^\dagger\dot U =(F_s^2/6\gamma){\mathbb G}'$ with
\begin{align}
\label{eq:g2withF}
&  {\mathbb G}' = \sum_\ka g'_\ka(Q_\ka ,P_\ka ) + g'_{c}\:,\nonumber\\
 &g'_\ka(Q,P) = \frac{1}{4}(P_\ka^2+Q_\ka^2-\mu_\ka')^2\nonumber\\
& +\frac{1}{2}\fp(P_\ka ^2-Q_\ka^2)-\frac{1}{4}\mu_\ka'^2\:.
\end{align}
Here, $g'_{c}$ is given by Eq.~(\ref{eq:coupling_RWA}) for $g_c$ in
which $V_{\ka \ka'} $ is replaced with $V'_{\ka \ka'} =2\ep_{\ka \ka'}
/F_s$ . The dimensionless time $\tau$, in which the
RWA dynamics is described by the equation $dA/d\tau =
-i(\lambda')^{-1}[A,{\mathbb G}']$,
is $\tau = (F_s/2\omega_F)t$. 

For ferromagnetic coupling in a periodic system of identical oscillators ($\ka\to \varkab, \mu'_\ka\to \mu'$) in the broken-symmetry state we have a minimum of ${\mathbb G}'$ at $Q_\varkab = \pm Q^{(0)\,\prime}$, $P_\varkab=0$, with
\begin{align}
\label{eq:barQwithF}
& Q^{(0)\,\prime}= (f-f_{\rm QPT})^{1/2}, \qquad f_{\rm QPT} = -\mu' -V'({\bf 0}),\nonumber\\ 
&\omega(\kb)= [2\fp+V'({\bf 0})-V'(\kb)]^{1/2}\nonumber\\
&\times [2f-2f_{\rm QPT} +V'({\bf 0})- V'(\kb)]^{1/2}.
\end{align}
Here, $V'(\kb)$ is given by Eq.~(\ref{eq:mode_frequencies}) for $V(\kb)$ with $V_{\varkab\varkab'}$ replaced by $V'_{\varkab\varkab'}$.

If $\mup$ is negative and $\mup + V'({\bf 0}) <0$,
Eq.~(\ref{eq:barQwithF}) leads to a critical value of the scaled driving force amplitude  $\fp= f_{\rm QPT} =  -\mup -V'({\bf 0})$ where $Q^{(0)\,\prime}=0$ and the gap in the excitation spectrum (\ref{eq:barQwithF}) disappears. The analysis of the case $f<f_{\rm QPT}$ is fully analogous to that for the case $\mu<\mu_{\rm QPT}$; in this case $Q^{(0)\,\prime}=0$. The results show explicitly that one can go through the quantum phase transition by either varying the driving frequency or the driving amplitude.


%

\end{document}